\documentclass[pra,twocolumn,superscriptaddress,aps]{revtex4-1}

\usepackage{stackrel}
\usepackage[binary-units]{siunitx}
\usepackage{color}

\usepackage[pdftex]{graphicx}
\usepackage{ulem}
\usepackage[hidelinks]{hyperref}
\usepackage{bm}
\usepackage{pifont}
\usepackage{anyfontsize}

\definecolor{mygreen}{rgb}{0,0.5,0} 

\newcommand{\be}{\begin{equation}}
\newcommand{\ee}{\end{equation}}
\newcommand{\bea}{\begin{eqnarray}}
\newcommand{\eea}{\end{eqnarray}}
\newcommand{\sdsbec}{SDSBEC}

\newcommand{\Ttwo}{\ensuremath{{T}_2} }
 \newcommand{\Tevol}{\ensuremath{T_{\rm evol}}}

\newcommand{\tProbe}{\tau}
\newcommand{\Trelax}{T_{\rm scat}}

\newcommand{\rate}[1]{\Gamma _#1}

\newcommand{\DensInt}[1]{I_{#1}}
\newcommand{\bc}{{\bf c}}

\renewcommand{\Tevol}{T}

\newcommand{\rTF}{r_{\rm TF}}
\newcommand{\rb}{\ensuremath{ ^{87}\mathrm{Rb} } }

\newcommand{\NA}{N}

\newcommand{\citeSI}{\cite}
\newcommand{\SM}{Methods}

\begin{document}

%\title{Ultracold ferromagnetic field sensor with energy resolution per bandwidth below $\hbar$}
\title{Single-domain Bose condensate magnetometer achieves energy resolution per bandwidth below $\hbar$}

\newcommand{\ICFOAddress}{ICFO-Institut de Ciencies Fotoniques, The Barcelona Institute of Science and Technology, 08860 Castelldefels (Barcelona), Spain} 
\newcommand{\ICREAAddress}{ICREA -- Instituci\'{o} Catalana de Recerca i Estudis Avan\c{c}ats, 08010 Barcelona, Spain} 
\newcommand{\IFAddress}{Instituto de F\'isica, Universidad Nacional Aut\'onoma de M\'exico, Apartado Postal 20-364, 01000 Cd. M\'exico, M\'exico}
\newcommand{\QCDAddress}{QCD Labs, QTF Centre of Excellence, Department of Applied Physics, Aalto University, Espoo, Finland}
\renewcommand{\QCDAddress}{Quantum Computing and Devices (QCD) Labs, Department of Applied Physics, Aalto University and Quantum Technology Finland (QTF) Centre of Excellence, FI-00076 Aalto, Finland}

% Use letters for affiliations, numbers to show equal authorship (if applicable) and to indicate the corresponding author
\author{Silvana Palacios Alvarez}
\affiliation{\ICFOAddress}
\author{Pau Gomez}
\affiliation{\ICFOAddress}
\author{Simon Coop}
\affiliation{\ICFOAddress}
\author{Roberto Zamora-Zamora}
\affiliation{\QCDAddress}
\author{Chiara Mazzinghi}
\affiliation{\ICFOAddress}
\author{Morgan W. Mitchell}
\affiliation{\ICFOAddress}
\affiliation{\ICREAAddress}

% Please give the surname of the lead author for the running footer
%\leadauthor{Palacios Alvarez} 

% Please include corresponding author, author contribution and author declaration information
%\authorcontributions{S.P.A. built the apparatus and performed the experiments with help from S.C., P.G. and C.M, analyzed the data and performed the TWA simulations. R.Z.Z. performed {Gross-Pitaevskii equation} simulations. M.W.M. supervised the work and developed the TWA analysis. Manuscript by S.P.A and M.W.M. with input from all authors.}
%\authordeclaration{The authors declare no competing interests.}
%%\equalauthors{\textsuperscript{1}A.O.(Author One) contributed equally to this work with A.T. (Author Two) (remove if not applicable).}
%\correspondingauthor{\textsuperscript{1}To whom correspondence should be addressed. E-mail: morgan.mitchell@icfo.eu}

% At least three keywords are required at submission. Please provide three to five keywords, separated by the pipe symbol.
%\keywords{Quantum sensing $|$ Magnetometry $|$ Bose-Einstein condensates } 

\begin{abstract}
We present a magnetic sensor with energy resolution per bandwidth $E_R < \hbar$. We show how a \textsuperscript{87}Rb single domain spinor Bose-Einstein condensate, detected by non-destructive Faraday-rotation probing, achieves single shot dc magnetic sensitivity of {\SI{72+-08}{\femto\tesla} measuring a volume $V=\SI{1091+-30}{\micro \meter\cubed}$ for \SI{3.5}{\second},} and thus $E_R =\SI{0.075+-0.016}{\hbar}$. We measure experimentally the condensate volume, spin coherence time, and readout noise, and use phase-space methods, backed by 3+1D mean-field simulations, to compute the spin noise. Contributions to the spin noise include one-body and three-body losses and shearing of the projection noise distribution, due to competition of ferromagnetic contact interactions and quadratic Zeeman shifts. Nonetheless, the fully-coherent nature of the single-domain, ultracold two-body interactions allows the system to escape the coherence vs.~density trade-off that imposes an energy resolution limit on traditional spin-precession sensors. We predict that other Bose-condensed alkalis, especially the antiferromagnetic $^{23}$Na, can further improve the energy resolution of this method.  

\end{abstract}

\maketitle
%\thispagestyle{firststyle}
%\ifthenelse{\boolean{shortarticle}}{\ifthenelse{\boolean{singlecolumn}}{\abscontentformatted}{\abscontent}}{}

% If your first paragraph (i.e. with the \dropcap) contains a list environment (quote, quotation, theorem, definition, enumerate, itemize...), the line after the list may have some extra indentation. If this is the case, add \parshape=0 to the end of the list environment.

{W}ell-known quantum limits profoundly, but not irremediably, constrain our knowledge of the physical world. Uncertainty relations %% \cite{HeisenbergZfP1927}
forbid precise, simultaneous knowledge of observables such as position and momentum. Parameter estimation limits, e.g., the standard quantum limit and ``Heisenberg limit,'' constrain our ability to measure transformations not subject to uncertainty relations, e.g., rotations \cite{BraginskiiSPU1975, HelstromJSP1969}. Both these classes of quantum limits admit trade-offs: uncertainty principles allow an observable to be precisely known if one foregoes knowledge of its conjugate observable, and parameter estimation limits allow better precision in exchange for a greater investment of resources, e.g., particle number.% \cite{GiovannettiS2004}.  

A qualitatively different sort of quantum limit is found in magnetic field sensing, where {well-studied sensor technologies are known to obey a quantum limit on the  \textit{energy resolution per bandwidth},
\be
\label{eq:ERDef}
E_R \equiv  \frac{\langle \delta B^2 \rangle  VT}{2\mu_0}. 
\ee
Here $\langle \delta B^2 \rangle$ is the mean squared error of the measurement, $V$ is the sensed volume, $T$ is the duration of the measurement, and $\mu_0$ is the vacuum permeability}\footnote{A related definition, scaling as $E_R \propto A^{3/2}$, $A \equiv $ active area, applies to planar sensors \cite{RobbesSAA2006, MitchellNJP2020}.}. %making $E_R$ useful for cross-technology comparisons \cite{RobbesSAA2006, YangPRAp2017, MitchellRMP2020}.

{A limit on $E_R$ constrains sensitivity when measuring the field in a given space-time region, without reference to any other physical observable, nor to any resource.  In contrast to other quantum sensing limits, this allows nothing to be traded for greater precision; it means that details of the field distribution are simply unmeasurable.  Known limits on $E_R$, derived from quantum statistical modeling, show that dc superconducting quantum interference devices (dc SQUIDs) \cite{TescheJLTP1977, KochPRL1980, RobbesSAA2006}, rubidium vapor magnetometers \cite{KominisN2003, Jimenez-MartinezBook2017} and immobilized spin-precession sensors, e.g., nitrogen-vacancy centers in diamond (NVD) \cite{ZhouPRX2020, MitchellNJP2020}, are all limited to $E_R \ge \alpha \hbar$, where $\hbar$ is the reduced Planck constant, and $\alpha$ is a number of order unity.  These limits though, are imposed by technology-specific mechanisms and no universal constraint is known that expands across other technologies %This limit is imposed by technology-specific mechanisms, e.g. magnetic dipole-dipole coupling in NVD;  no universal constraint, such as uncertainty relations or quantum speed limits, is known to generalize this limit to other technologies 
	\cite{MitchellRMP2020}.

A variety of exotic sensing techniques, including noble-gas {spin-precession sensors} \cite{NewburyPRA1993, KochEPJD2015, KochEPJD2015b}, levitated ferromagnets \cite{JacksonKimballPRL2016, VinanteARX2019} and dissipationless superconducting devices \cite{LuomahaaraNC2014, BalNC2012, DanilinNPJQI2018} have been proposed to achieve $E_R < \hbar$ by evading specific relaxation mechanisms \cite{MitchellRMP2020}.  If  $E_R < \hbar$ can be achieved, it will break an impasse that has held since the early 1980s, when $E_R \approx \hbar$ was reached in dc SQUID  sensors \cite{KochPRL1980, Cromar1981}.  In addition to resolving the question of whether $E_R \ge \hbar$ is universal, achieving $E_R < \hbar$ would open horizons in condensed matter physics \cite{YangPRAp2017} and neuroscience \cite{BotoNI2017}. For example: to enable single-shot discrimination of brain events, a magnetometer would need $\delta B \sim \SI{1}{\femto\tesla}$ sensitivity to $T \sim \SI{10}{\milli\second}$ events when measuring in $V \sim (\SI{3}{\milli\meter})^3$ volumes \cite{MacgregorNC2012, PrattSPIE2021}, or $E_R \sim 1 \hbar$. 

Here we study an exotic magnetometer technology,} the single-domain spinor Bose-Einstein condensate (SDSBEC), that freezes-out relaxation pathways due to collisions, dipolar interactions, and also spin diffusion \cite{Vengalattore2007} and domain formation \cite{LeePRA2016, Jimenez-GarciaNC2019}, which occur in unconfined condensates. With a \textsuperscript{87}Rb SDSBEC, we find $E_R=\SI{0.075+-0.016}{\hbar}$, far beyond what is  possible, even in principle, with established technologies \cite{GroszBook2016, MitchellRMP2020}.  Our results demonstrate the possibility of $E_R \ll \hbar$ sensors, and motivate the study of other exotic sensor types.

\begin{figure*}[t]
\centering
\includegraphics[width =0.95 \textwidth, trim=0cm 0cm 0cm 0cm, clip=true]{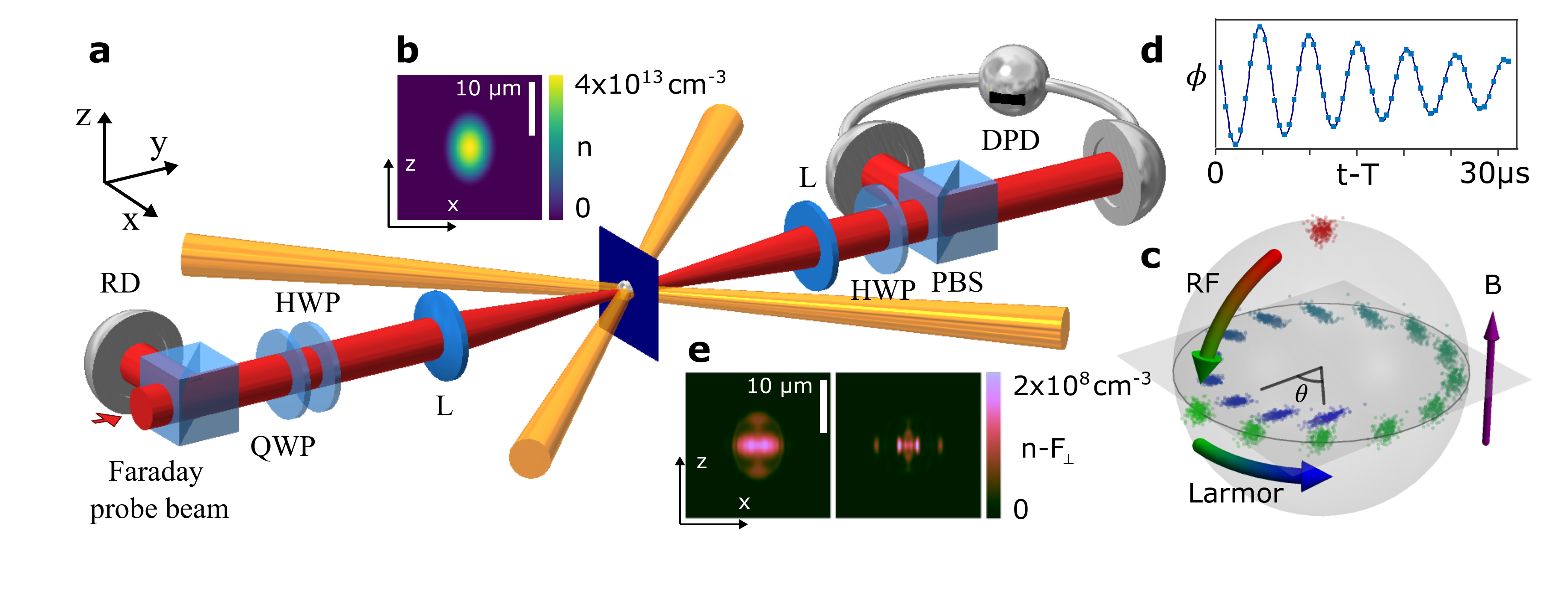}%
\caption{Single-domain spinor Bose-Einstein condensate (\sdsbec)~magnetic field sensor. (a) Experimental schematic: crossed, far-off-resonance beams (orange) are used to produce and hold a spinor condensate in a spherical {optical dipole trap}.  A near-resonance probe beam (red) is used to make non-destructive Faraday rotation measurements of the on-axis ($y$) component of the collective spin ${\bf F}$. A reference detector (RD) measures the number of input photons, quarter- (QWP) and half-wave (HWP) plates are used to set the polarization before a lens (L) focuses the probe onto the atomic cloud. The transmitted light is analyzed for polarization rotation using a second HWP,  polarization beamsplitter (PBS) and differential photodetector (DPD).  (b)
Computed density $n$  of the prepared SBEC in the $x$--$z$ plane (dark square in schematic).  (c) Evolution of the collective spin statistical distribution during the sensing protocol (not to scale): the atoms are spin polarized parallel to the field direction, with the collective spin ${\bf F}$ statistically distributed as shown by red dots, limited by spin projection noise and atom number uncertainty.  Spins are then tipped by a radio-frequency pulse to be orthogonal to the field (B), shown by green dots. During a free-precession time $T$ the collective spin precesses by an angle $\theta=\gamma B T$, while also diminishing in magnitude and experiencing shearing of the statistical distribution (green-blue progression).  (d) Readout: during the final few precession cycles the spin component $F_y$ is detected by Faraday rotation.  Measurements of optical polarization rotation angle $\phi$ versus time $t$ (points) are fit with a free-induction waveform (line) to infer spin-rotation angle $\theta$ at readout time $T$. (e)  Spatial distribution of the  polarization defect density $n -F_\perp$ at $T=\SI{1}{\second}$, where  $F_\perp$  is the transverse polarization density, obtained from 3+1D {Gross-Pitaevskii equation} simulations for the experimental trap conditions and $q/h=\SI{0.5}{Hz}$ (left) and $q/h=\SI{0}{Hz}$ (right). Scale as in (b).  The very small observed spin defect implies a small upper bound to spin noise from ferromagnetism-driven spin segregation, and justifies the use of the single-mode approximation to compute quantum noise dynamics. 
}
 \label{fig:SetupSequence}
\end{figure*}

%We consider a {\sdsbec} that  operates as a spin-precession sensor (SPS) in its hyperfine ground state. 
  
 To understand how the {\sdsbec} evades the $\hbar$ limit, it is instructive to first show why other {spin-precession sensors}, which include NVD and alkali vapors, obey such a limit. The principle of operation of a {spin-precession sensor} is represented in \autoref{fig:SetupSequence}{\bf c}: An ensemble of $N$ atoms is first initialized with its net spin ${\bf F}$ along the magnetic field ${\bf B}$ to be measured. The spin is then tipped by a radio-frequency pulse, making ${\bf F}$ orthogonal to ${\bf B}$. The spins are allowed to precess for a time $T$ before the resulting precession angle $\theta = \gamma B T$ is detected, where $\gamma$ is the gyromagnetic ratio of the atomic species and $B = |{\bf B}|$ is the magnitude of the field.  The resulting energy resolution per bandwidth is
\be
\label{eq:deltaBsqT}
E_R 
%\propto \langle \delta B^2 \rangle V T 
=   \frac{V \langle \delta \theta^2 \rangle_F }{2 \mu_0 \gamma^2 T}  +  \frac{V \langle \delta \theta^2 \rangle_{\rm RO} }{2 \mu_0 \gamma^2 T}, 
\ee
where $\langle \delta \theta^2 \rangle_F$ and $\langle \delta \theta^2 \rangle_{\rm RO}$ are the angular variance due to intrinsic uncertainty of ${\bf F}$ and readout noise, respectively.

%$\langle \delta \theta^2 \rangle_{\rm RO}$ 
Readout noise can in principle be arbitrarily reduced using projective measurement, so we focus on the intrinsic spin noise. This scales as $\langle \delta \theta^2 \rangle_F \propto N^{-1}$ and is minimized at the optimal readout time $T \approx T_2/2$, where $T_2$ is the transverse relaxation time. The quantum noise contribution to \autoref{eq:deltaBsqT} thus scales as $1/(n T_2)$, where $n = N/V$ is the number density of spins.  In ordinary spin systems, the relaxation rate $1/T_2$ will grow proportionally to $n$ due to two-body decoherence processes, e.g., spin-destruction collisions in vapors \cite{Jimenez-MartinezBook2017} or magnetic dipole-dipole coupling in NVD \cite{MitchellRMP2020, ZhouPRX2020}.  This {density-coherence trade-off} ensures that $E_R$ has a finite lower bound  (see  \SM, \autoref{sec:SPSERL}). 
 
{To circumvent this  limit, we implement a spin-precession sensor with a SDSBEC. This ultra-cold sensor} differs from the above in three important ways.  First, because it is so cold, inelastic two-body  interactions, including both short-range hyperfine-changing collisions and long-range dipole-dipole interactions, are energetically forbidden for a sensor operating in the ground hyperfine state \cite{MiesJRNIST1996}.  Second, because of quantum degeneracy, the elastic two-body interactions (spin-independent and spin-dependent contact interactions) produce a coherent dynamics that does not raise the entropy of the many-body spin state \cite{PalaciosNJP2018}.  Third, in the single-domain regime, these coherent dynamics cannot reduce the net polarization through domain formation, as happens in extended SBECs \cite{Sadler2006}.  As we will show, $1/T_2$ then contains no contribution $\propto n$, and we escape the density-coherence  trade-off. %that limits $E_R$ in other \ntext{spin-precession sensors}.

% In the single-domain condition these dynamics preserve full polarization of the spin ensemble   This allows $1/T_2$ to scale As a consequence, $1/T_2$ doesn't have the same scaling with n allowing a different 1/T2n,
%the spin quantum noise contribution, to approach zero as we will show. 
%
%This suggests that it is possible to avoid all two-body relaxation, and thus alter the scaling of $1/T_2$ with $n$, allowing $1/(n T_2)$, and thus the spin quantum noise contribution, to approach zero. 
% 
%In SBECs, in contrast, the entropy is much less than one bit per atom.  Two-body dissipation mechanisms: dipolar-relaxation from hyperfine-changing collisions and dipole-dipole interactions, are energetically forbidden \cite{MiesJRNIST1996}.  Additionally, the ferromagnetic interaction  is coherent and in this single domain geometry no domain formation is allowed \cite{PalaciosNJP2018}. This lack of entropy required to decohere the spin degrees of freedom through the interaction with other degrees of freedom makes the SDSBEC able to operate as spin-precession sensors  free of two-body relaxation and, in turn, escape the $\hbar$-limit  imposed in other SPS systems. 

To understand the SDSBEC sensitivity\footnote{A direct measurement of the sensor's equivalent magnetic noise could in principle be made by placing the magnetometer in a shielded environment with magnetic noise below that of the sensor. To our knowledge, shielding at the required level, $\sim \SI{50}{\femto\tesla\per\sqrt\hertz}$ at sub-\SI{}{\hertz} frequencies, has never been implemented in a cold-atom experiment, and appears intrinsically challenging.  As described below, the single-shot, optimized SDSBEC is sensitive to frequencies below $f = 1/T \approx \SI{0.29}{\hertz}$, while multi-shot measurements would be still slower.  At these low frequencies, magnetic shielding is limited by the innermost shield's thermal magnetization noise, with power spectral density $\propto 1/f$ and typical values $\langle \delta B^2 \rangle T = f^{-1} \SI{120}{\femto\tesla\squared}$ \cite{KornackAPL2007}. For this reason, we base our sensitivity estimates on a combination of measured readout noise and calculations of the quantum noise dynamics in the SBEC using measured parameters. Due to the very clean nature of the BEC system, such calculations have proven reliable in other contexts \cite{HeFP2012}.}, we compute $\langle \delta \theta^2 \rangle_F$, including quantum statistical effects due to collisional interactions, which can importantly modify the spin distribution from its mean-field behavior \cite{LuckeS2011}. 
% we compute the resulting evolution of the spin distribution during the evolution time $T$ 
We employ the truncated Wigner approximation (TWA) \cite{SteelPRA1998, OpanchukJMP2013}, previously applied to study spatial coherence in BECs \cite{SinatraJPB2002}.   In the single-mode approximation (SMA), the quantum field describing the condensate factorizes into a spatial distribution $\phi_N({\bf r})$ and a spinor field operator $\chi$ describing all atoms in the condensate. ${\bf \chi} \equiv (\hat{a}_{+1}, \hat{a}_0 , \hat{a}_{-1})^T$ where $\hat{a}_m$  are bosonic annihilation operators, such that $N \equiv {\bf \chi}^\dagger \cdot {\bf \chi}$ is the atomic number operator.  $\phi_N({\bf r})$ is the ground-state solution to the spin-independent part of the Hamiltonian in the Thomas-Fermi approximation and with $N$ atoms.  We normalize $\phi_N$ such that $\DensInt{2} = 1$, where $\DensInt{d} \equiv \int d^3 \vec{r} \, |\phi_N(\vec{r})|^d$.  

The spinor field $\chi$ evolves under the SMA Hamiltonian \cite{StamperRMP2013} 
\begin{eqnarray}
\label{eq:HSMA}
H_{\rm SMA} &=&   
  \frac{g}{2} \chi^\dagger  {\bf f} \chi \cdot \chi^\dagger {\bf f} \chi +  q  \chi^\dagger f_z^2 \chi , %p  \chi^\dagger F_z \chi + 
\end{eqnarray}
where $g \equiv {g_2}{} I_4 \propto N^{-3/5}$ describes the spin-dependent interaction strength and the $q$ term describes the quadratic Zeeman shift, including contributions from the external field and from microwave or optical fields.  The combined action of the $q$ and $g$ terms induces a shearing of the condensate's spin noise distribution from its initial coherent-state distribution.  Losses occur at rate $dN/dt = -\rate{1} N - \rate{3} N^{9/5}$, where  $\rate{1}$ describes the rate of collisions with background gas and $\rate{3}$ is proportional to the three-body loss cross section. % both reduce the number of atoms and introduce spin noise, as illustrated in \autoref{fig:SimulationResults}, insets.
The evolution of the many-body spin state $\rho$ is described by the master equation $d\rho/dt = [H_{\rm SMA},\rho]/({i\hbar}) + {\cal L}[\rho]$ where ${\cal L}[\rho]$ is the Liouvillian
\be
{\cal L}[\rho] = \sum_{l} \kappa_l  \left( 2 \hat{O}_l \rho \hat{O}_l^\dagger - \rho \hat{O}_l^\dagger \hat{O}_l - \hat{O}_l^\dagger \hat{O}_l \rho  \right),
\ee
and the ``jump operators'' $\hat{O}_l$, with associated rates $\kappa_l$, describe the various loss processes (see \SM, \autoref{sec:modeshape} and \autoref{sec:qnoise}).

\begin{figure}[t]
 \includegraphics[width = \columnwidth, trim=0mm 0mm 0mm 0mm,clip=true]{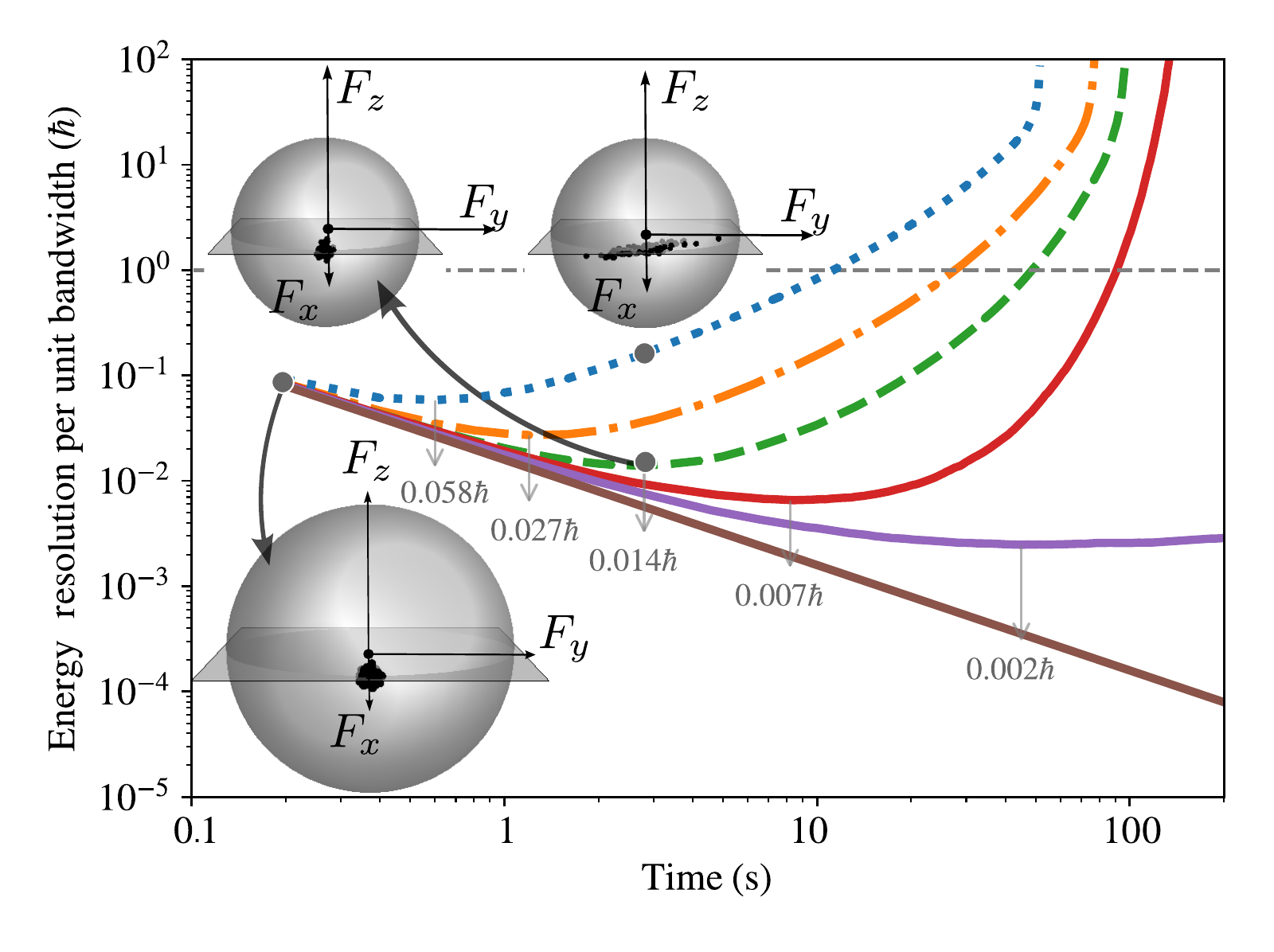}
 \caption{Spin noise contribution to $E_R$ of the SDSBEC sensor, from TWA simulations with measured  trap parameters including condensate volume $V$ and one- and three-body decay rates $\rate{1}$ and $\rate{3}$, respectively.  Blue, orange, green, and red curves show $E_R$ for $q/h = 0.30, 0.12, 0.05$ and \SI{0}{Hz}, respectively.  To separate different effects, we show also conditions $\rate{1}= q/h={0}{}$ (violet), and $\rate{1} = \rate{3} = q/h={0}{}$ (brown). Spheres represent the $(F_x, F_y , F_z)$ phase space at time \SI{0}{\second} (bottom) and at \SI{3}{\second} with $q/h=$ \SIlist{0.3;0}{\hertz} (top left, top right); sphere radius is equal to the number of remaining atoms, points sample the rotating-frame Wigner distribution. For ease of visualization, dispersion of the Wigner distribution is magnified by a factor of 10.}
 \label{fig:SimulationResults}
\end{figure}

 \autoref{fig:SimulationResults}  shows the evolution of the spin noise contribution to $E_R$ over time as computed by TWA. For a given trapping potential and finite $\rate{1}$, $q$, and/or $\rate{3}$, the energy resolution shows a global minimum with $T$. To understand the in-principle limits of this $T$-optimized noise level, we note the following: 
 %(see \SM, \autoref{sec:scaling}): 
 1) $\rate{1}$  can in principle be arbitrarily reduced through improved vacuum conditions, while $q$ can also be made arbitrarily small by compensating the contribution of the external field  with microwave or optical dressing, leaving $\Gamma_3$ as the sole factor to introduce spin noise.  2) The noise effects of $\Gamma_3$, which are a strong function of density, can also be made arbitrarily small, by increasing $\rTF$ and $N$ to give a large, low-density condensate. 3) The corresponding increase in $V$ is more than offset by the increase in $T_2$, such that $E_R \propto V/T_2$ tends toward zero. 4) At the same time, the SMA and TWA approximations become more accurate in this limit. We conclude that a low-density SBEC in a loose trapping potential can operate deep in the single-mode regime, suffer small three-body losses, and achieve $E_R \ll \hbar$.  %The SDSBEC sensor thus has unlimited energy resolution. \mtext{weaken}

We now show that a SDSBEC magnetometer can in practice operate with $E_R$ well below $\hbar$.   The experimental configuration is illustrated in \autoref{fig:SetupSequence}{\bf a}, and described in detail in  Palacios \textit{et al.} \cite{PalaciosNJP2018}. In brief, a  pure condensate of \rb atoms in the $F=1$ manifold with an initial atom number $N_0= \SI{6.8+-.5e4}{}$ is produced by forced evaporation in a crossed-beam optical dipole trap. The condensate is initialized fully polarized along ${\bf B}$ by evaporation in the presence of a magnetic gradient, tipped by a radio-frequency pulse to be orthogonal to ${\bf B}$, then allowed to precess for a time $T$ before read-out, as depicted in \autoref{fig:SetupSequence}{\bf c}. Probe light tuned \SI{258}{\mega\hertz} to the red of the $F=1 \rightarrow F'=0$ transition of the D$_2$ line is used for non-destructive Faraday rotation measurement of the collective spin of the condensate (\autoref{fig:SetupSequence}{\bf {d}}). 

Atom number is measured by time-of-flight absorption imaging. From atom-number decay we observe $\rate{1}=\SI{8.6+-3.1e-2}{\per\second}$ and $\rate{3}=\SI{1.0+-0.6e-5}{atom^{-4/5}\second^{-1}}$ one-body and three body collision rates, respectively. The very small three body loss rate allows us to approximate atomic losses as exponentially decaying with lifetime \SI{7.1+-.2}{s}.  In this approximation the resulting rate $dN/dt$ never differs by more  than $4\%$ from the numerical solution when both $\rate{1}$ and $\rate{3}$ are included. The coherence time is found to be equal to the atomic lifetime in the trap and therefore  $\Ttwo=\SI{7.1+-.2}{s}$. % \mtext{need to explain the relation between $\rate{1}$ and $T_2$}

 The curvature of the trapping potential is determined from the measured SBEC oscillation frequencies. We find $\omega_1/2\pi =$  \SI{67.2 +- 1.0}{Hz}, $\omega_2/2\pi=$  \SI{89.0 +- 0.7}{Hz} and $\omega_3/2\pi=$ \SI{97.6 +- 0.9}{Hz}, where the subscripts index the principal axes of the trap.  For our number of atoms $\NA=\SI{6.8+-.5e4}{atoms}$ these correspond to Thomas-Fermi radii $r_{\rm TF}^{(1,2,3)}=\SI{7.0+-0.1}{\micro m}$, $\SI{6.2+-0.09}{\micro m}$  and $ \SI{6.0+-0.09}{\micro m}$ in the {Thomas-Fermi approximation}  (see \SM, \autoref{sec:modeshape}). This parabolic geometry defines the volume containing the entire condensate $V\equiv 4 \pi r_{\rm TF}^{(1)}r_{\rm TF}^{(2)}r_{\rm TF}^{(3)}/3$ $=\SI{1091+-30}{\micro \meter\cubed}$.
 
  As shown in \autoref{fig:SetupSequence}{\bf {d}}, measurements of the spin precession can be taken over several precession cycles with little damage to the polarization, allowing the precession angle to be estimated with readout noise  $\langle \delta \hat{\theta} ^2\rangle_ {\rm RO}=\SI{1.08+-0.24e-4}{rad^2}$ at the time of optimal readout $T=\Ttwo/2$ (see \SM, \autoref{sec:Readout}). We note that $\langle \delta \theta^2 \rangle_{\rm RO}$ could be further reduced through improved probe-atom coupling {and/or squeezed light \cite{PredojevicPRA2008, TroullinouPRL2021}}.  
  
Combining the above we have  volume $V=\SI{1091+-30}{\micro \meter\cubed}$, readout noise  $\langle \delta \hat{\theta} ^2\rangle_ {\rm RO}=\SI{1.08+-0.24e-4}{rad^2}$ and spin quantum noise  $\langle \delta \theta^2 \rangle_F=\SI{1.46+-1e-5}{rad^2}$. For an optimum read out time of $T=\SI{3.5}{s}$, these give a magnetic sensitivity of \SI{72+-08}{\femto\tesla} and $E_R=\SI{0.075+-0.016}{\hbar}$ (see \SM, \autoref{sec:dutycycle}). This is a factor of 17 better than any previously reported value \cite{AwschalomAPL1988, WakaiAPL1988, Vengalattore2007} and well beyond the level $E_R \approx \hbar$ that constrains the most advanced existing technologies.

%  In our experiment we have measured volume $V=\SI{1091+-30}{\micro \meter\cubed}$, readout noise  $\langle \delta \hat{\theta} ^2\rangle_ {\rm RO}=\SI{1.08+-0.24e-4}{rad^2}$ and estimated $\langle \delta \theta^2 \rangle_F=\SI{1.46+-1e-5}{rad^2}$ which for an optimum read out time of \SI{3.5}{s} In the following we detail how we obtain these numbers and discuss the SBEC-specific effects, including spin domain formation and quantum noise amplification (anti-squeezing of the intrinsic spin noise), that could, in other scenarios, prevent a SBEC from operating as a dissipation free-SPS. 

 In applying the TWA, we assumed the validity of the single-mode approximation. To check this, we integrate in time the three-dimensional Gross-Pitaevskii equation (see \SM, \autoref{sec:descriptioncondensate}) on a graphical processing unit, as described in \cite{VillasenorPRA2014, Zamora2018}.  Spatially-resolved polarization densities are shown in \autoref{fig:SetupSequence}{\bf {b}} and {\bf {e}} and indicate fractional polarization defects at the $10^{-5}$ level. The defect $\NA - F_\perp$ of the condensate as a whole is of order 1 atom. By vector addition, the contribution to the variance of the azimuth spin  component $F_\theta$ is then no larger than the projection noise $\langle \delta F_\theta^2 \rangle_{\rm PN} = \NA/2$, and could be far smaller. These mean-field results, together with coherence measurements reported in \cite{PalaciosNJP2018}, give a quantitative justification for the use of the single-mode approximation. % \cite{Ho1998, Law1998SMA, Yi2002}.

\begin{figure}[t]
 \includegraphics[width = \columnwidth, trim=0mm 0mm 0mm 0mm,clip=true]{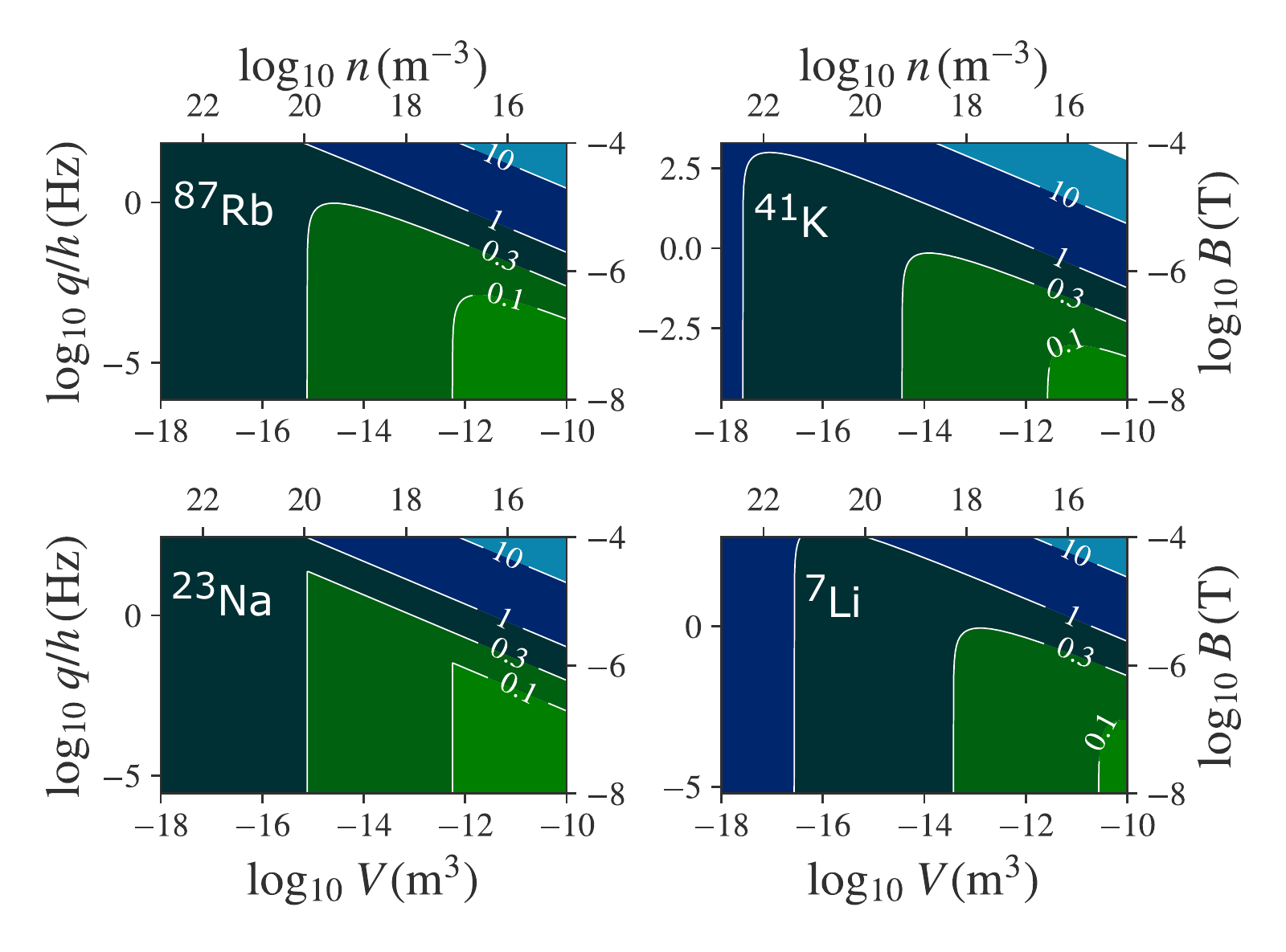}
 \caption{Comparison of alkali atoms with $F=1$ ground states as SDSBEC sensors.  The SMA will be valid for $\rTF/\lambda\ll 1$ and $\rTF/\xi_s \ll 1$, where $\lambda$ is the threshold wavelength for spin-wave amplification and $\xi_s$ is the spin healing length.  Graphs show $\max(\rTF/\lambda,\rTF/\xi_s)$ versus volume $V$ (shown also as number density $n$) and $q$ (shown also as field strength $B$) for $\NA = \SI{6.8e4}{}$. }
 \label{fig:stability}
\end{figure}

We extend the analysis to other $F=1$ alkali species and find that some could perform still better than the \rb system studied here.  Two considerations are relevant here.  First, we note the conditions for single-mode dynamics:  $\rTF/\xi_s \ll 1$ and  $\rTF/\lambda \ll 1$, where  $\xi_s$ is the spin-healing length \cite{StamperRMP2013} and  $\lambda$ is the threshold wavelength for spin-wave amplification \cite{makela2011} (see \SM, \autoref{sec:SMAconditions}).  In \autoref{fig:stability} we show $\max( \rTF/\xi_s , \rTF/\lambda)$ versus $V$ and $q$, and note that \rb and $^{23}$Na remain single-domain for smaller volumes and for stronger fields than do $^{7}$Li and $^{41}$K.  We note also that the dynamical condition $\rTF/\lambda \ll 1$ favors anti-ferromagnetic interactions, %i.e. $g_2 > 0$, 
giving $^{23}$Na a marked advantage by this criterion.  The second consideration concerns the three-body recombination rate \cite{Fedichev1996} $\rate{3}\propto \hbar a_{0}^4/M$, where $a_{0}$ is the s-wave scattering length for the channel of total spin zero. Relative to $^{87}$Rb, this rate in $^7$Li, $^{23}$Na and $^{41}$K is a factor 25, 4 and 2 smaller, respectively, suggesting an advantage for these species when limited by three-body losses.

In conclusion, we have shown that an appropriately confined, quantum degenerate Bose gas, i.e., a single-domain spinor Bose-Einstein condensate (SDSBEC), has a {qualitative} advantage over the best existing magnetic sensors as regards temporal, spatial, and field resolution, as summarized in the energy resolution per bandwidth $E_R$.  Whereas the best-developed approaches to superconducting, hot vapor, and color center magnetometers are limited to $E_R \gtrsim \hbar$, the SDSBEC, which retains a strong global response to an external field, while freezing out internal interactions that would otherwise produce depolarization, can operate with $E_R$  far below $\hbar$.  With a $^{87}$Rb SDSBEC, we have demonstrated  $E_R = 0.075(16) \hbar$, a factor of 17 improvement over the best previously reported \cite{AwschalomAPL1988, WakaiAPL1988, Vengalattore2007} and well beyond the level that limits todays most advanced magnetic sensors. $E_R$ in the demonstrated \textsuperscript{87}Rb system could be reduced with better light-atom coupling. Other alkali SBECs could also achieve smaller values for $E_R$. The results show the promise of a new generation of proposed sensors, including noble-gas magnetometers \cite{NewburyPRA1993, KochEPJD2015, KochEPJD2015b}, levitated ferromagnets \cite{JacksonKimballPRL2016, VinanteARX2019}, and dissipationless superconducting devices \cite{LuomahaaraNC2014, BalNC2012, DanilinNPJQI2018}, that operate by similar principles. 

\renewcommand{\thesection}{M\arabic{section}}  
\renewcommand{\thetable}{M\arabic{table}}  
\renewcommand{\thefigure}{M\arabic{figure}}
\renewcommand{\theequation}{M\arabic{equation}} 
\setcounter{equation}{0} 

\newcommand{\showmethods}
{\appendix
\section{Energy resolution limit for Markovian spin systems}\label{sec:SPSERL}We describe an ensemble of $\NA$ spin-$F$ atoms by the collective spin operator ${\bf F}$, i.e., the sum of the vector spin operators for the individual atoms. ${\bf F}$ is initialized in a fully-polarized state orthogonal to the magnetic field ${\bf B}$.  The spin angle precesses at a rate $\dot{\theta} = \gamma B$ where $\gamma$ is the gyromagnetic ratio. It is convenient to work with spin components in a frame rotating at the nominal Larmor frequency, such that a small change in angle can be expressed as $\delta \theta = \delta F_\theta / F_\perp$, where $F_\theta$ is the azimuthal component and $F_\perp$ is the ``lever arm'' or spin component orthogonal to the axis of rotation, and thus orthogonal to ${\bf B}$. If a measurement of $F_\theta$ is made at time $T$ to infer $\theta$ and thus $B$,  the equivalent magnetic noise is $\langle \delta B^2 \rangle = \langle \delta \theta^2 \rangle /(\gamma^2 T^2)$, by propagation of error.  If $F_\perp$ experiences Markovian relaxation, then $F_\perp$ at the time of measurement is $F_\perp(T) = F \NA \exp[-T/T_2]$, where $T_2$ is the transverse relaxation time and $F \NA$ is the initial, full polarization.  The initial, fully-polarized state has azimuthal spin noise $\langle \delta F_\theta^2 \rangle = F \NA /2$, i.e., the standard quantum limit.  If $\NA$ does not decrease during the evolution (as is the case for color center and vapor phase ensembles), this describes a minimum noise for $F_\theta$ during the evolution.  We thus find $\langle \delta B^2 \rangle T \ge \exp[2T/T_2]/(2 \gamma^2 T F \NA)$.  Choosing $T$ to minimize the r.h.s. of this inequality, we find $T = T_2/2$ and thus $\langle \delta B^2 \rangle T \ge \exp[1]/(2 \gamma^2 T_2 F \NA)$.  Including the sensor volume $V$, the energy resolution is lower-bounded by $E_R \ge  \exp[1]/(4 \mu_0 \gamma^2 F T_2 n)$, where $n=N/V$ is the number density.  

Writing the relaxation rate as $1/T_2 = A_1 n^0 + A_2 n^1 + \ldots$,  $d$-body interactions contribute to the $A_{d}$ term. When $A_2$ is non-zero, $E_R \propto A_1 n^{-1} + A_2 n^0 + \ldots $ is manifestly lower-bounded.  First principle calculations for immobilized {spin-precession sensors} \cite{MitchellNJP2020}, and models including measured spin-relaxation rates for optimized Rb vapor magnetometers \cite{Jimenez-MartinezBook2017} show that these lower bounds are within a factor of two of $E_R = \hbar$.

\section{Description of the condensate}
\label{sec:descriptioncondensate}
A $F=1$ spinor condensate with weak collisional interactions is well described by a three-component field $\psi_\alpha(\bf r)$ evolving under the hamiltonian
\begin{eqnarray}
\label{eq:H}
H & = & H_{\rm SI} + H_{\rm SD}  
\end{eqnarray}
where $H_{\rm SI}$ and $H_{\rm SD}$ are the spin-independent and spin-dependent parts, respectively.  Summing over repeated indices, and omitting position dependence for clarity, these are
\begin{eqnarray}
\label{eq:HSI}
H_{\rm SI} &=& \int d^3r \, \left( \psi_\alpha^\dagger [- \frac{\hbar^2 \nabla^2}{2 M} + U ] \psi_\alpha + \frac{g_1}{2} \psi^\dagger_\alpha \psi^\dagger_\beta \psi_\beta \psi_\alpha \right) \hspace{5mm} 
 \\ 
 \label{eq:HSD}
H_{\rm SD} &=& \int d^3r \, \frac{g_2}{2}  \psi_\alpha^\dagger (f_\eta)_{\alpha\beta} \psi_\beta \psi_\gamma ^\dagger (f_\eta)_{\gamma\delta} \psi_\delta  + p  \psi_\alpha^\dagger (f_{z})_{\alpha\beta} \psi_\beta 
\nonumber \\ & & +   q  \psi_\alpha^\dagger (f_z f_z)_{\alpha\beta} \psi_\beta. 
\end{eqnarray}
Here $f_\eta$ is the matrix representing the single-atom spin projection operator onto the axis $\eta$. 
In $H_{\rm SD}$, the terms are ferromagnetic interaction, linear Zeeman and quadratic Zeeman energies, respectively,  $p =\hbar \gamma B$, and $q =(\hbar \gamma B)^2/E_{\rm hf}$, where $B$ is the field strength and $E_{\rm hf}$ the hyperfine splitting energy. S-wave scattering contributes the state-independent and state-dependent contact interactions, characterized by $g_1 \equiv 4 \pi \hbar^2 (a_{0} + 2 a_{2})/(3 M)$ and  $g_2 \equiv 4 \pi \hbar^2 (a_{2} - a_{0})/(3 M)$, respectively. Here $M$ is the atomic mass and $a_{0}, a_{2}$ are the s-wave scattering lengths for the channels of total spin 0 and 2, respectively \citeSI{Kawaguchi}. We neglect the magnetic dipole-dipole interaction, which in \rb is orders of magnitude weaker than the contact interactions, and vanishes identically for a single-mode spherical distribution.

\section{Mode shape, interaction strengths, and jump operators}
\label{sec:modeshape}

In the {Thomas-Fermi approximation} \citeSI{Soding1999}, a pure condensate in a spherical harmonic potential has the mode function
\be
\label{eq:TFProfile}
|\phi(r)|^2 = 
\frac{15}{8 \pi \rTF^3} (1-\frac{r^2}{r_{\rm TF}^2}) 
\ee 
for $r\le \rTF$ and zero otherwise, where $r$ is the radial coordinate, $r_{\rm TF}= \left[15 g_1 \NA /(4 \pi M\omega^2) \right]^{1/5}$ is the Thomas-Fermi radius, and $\omega$ is the trap angular frequency.  Because $\rTF \propto N^{1/5}$,
the integrals $I_d$ that determine the effective strength of two- and three-body interactions are $I_4 \propto \NA^{-3/5}$ and $I_6 \propto \NA^{-6/5}$, respectively.  The rate of three-body collisions can then be written $\rate{3} \NA^{9/5} \propto I_6 N^3$, such that atom losses are described by 
\begin{equation}
 \frac{dN}{dt}=-\rate{1} N-\rate{3}N^{9/5}.
 \label{eq:dNdt}
\end{equation}
We note that in this model losses are independent of internal state. While this is well established for one-body losses, for three-body losses the state dependence is, to our knowledge, unknown.  Two-body losses due to magnetic dipole-dipole scattering and spin-orbit interaction in second order \cite{MiesJRNIST1996} are energetically forbidden in the low-field scenario of interest here.

We use a set of jump operators that reproduces \autoref{eq:dNdt} while also respecting the symmetry of the loss process:  One-body losses are described by $\hat{O}_{m}^{(1b)} = \hat{a}_{m}$, $m \in \{-1,0,1\}$, where $\hat{a}_m$ annihilates an atom in internal state $m$, with strengths $\kappa_{m}^{(1b)}  = \rate{1}/2$, while three-body losses are described by $\hat{O}_{mno}^{(3b)} = \NA^{-3/5} \hat{a}_{m}\hat{a}_{n}\hat{a}_{o}$, $m,n,o \in \{-1,0,1\}$, where $\NA \equiv (\hat{a}^\dagger_{-1} \hat{a}_{-1} + \hat{a}^\dagger_0 \hat{a}_0 + \hat{a}^\dagger_{+1} \hat{a}_{+1})$ with strengths $\kappa_{mno}^{(3b)} = 5 \Gamma_3/24$.

\section{Quantum noise evolution}
\label{sec:qnoise}

We use the truncated Wigner appoximation (TWA) \citeSI{JackPRL2002,SteelPRA1998,HeFP2012}  to compute the evolution of the spin distribution arising from  the master equation    $d\rho/dt = [H_{\rm SMA},\rho]/({i\hbar}) + {\cal L}[\rho]$. Our treatment follows that of Opanchuk \citeSI{OpanchukEPL2012}, restricted to a single spatial mode.  In the TWA,  the Wigner-Moyal equation describing the time evolution of the Wigner distribution is truncated at second order, such that an initially positive Wigner distribution remains positive, and the Wigner-Moyal equation becomes a Fokker-Planck equation\footnote{{The approximation is believed valid for non-critical systems in which each simulated mode contains on average many particles \cite{SinatraJPB2002, OpanchukThesis2014}. This condition is very well satisfied here.}}.  The Fokker-Planck equation describes the evolving probability distribution of a particle undergoing brownian motion, and as such can be described by a stochastic differential equation that is straightforward to integrate numerically. 

We identify a complex-valued vector $\bc = (c_{+1}, c_0, c_{-1})^T$ with the spinor field $\chi$, and c-number functions ${O}_{m}^{(1b)} = c_{m}$, $m \in \{-1,0,1\}$,  ${O}_{mno}^{(3b)} = |{\bf c}|^{-6/5} c_{m}c_{n}c_{o}$, $m,n,o \in \{-1,0,1\}$, with the jump operators $\hat{O}_{m}^{(1b)}$ and  $\hat{O}_{mno}^{(3b)}$, respectively.  To account for the uncertainty of the initial state, a collection of starting points are chosen with values $
\bc_i = \bc_0 + (z_{-1},z_0, z_{+1})^T/\sqrt{2}$, where $\bc_0 = (1, \sqrt{2}, 1)^T/2$ is the initial, fully $F_x$-polarized state,  $z_m = x_m + i y_m$ and $x_m$, $y_m$ are zero-mean unit-variance gaussian random variables. For the simulations shown in \autoref{fig:SimulationResults} we used 5000 starting points.

Each initial point evolves by the  (It\^{o})  {stochastic differential equation}
\begin{equation}\label{eq:GPE}
 {d} c_m= \left[ \frac{1}{i\hbar}\frac{\partial H}{\partial c_m^* } - \sum_{l} \kappa_{l}  \frac{\partial O_{l}^*}{\partial{c^*_m}}  O_{l} 
%\frac{\partial O_{l}^*}{\partial c_j^*}  
\right] d t
% \nonumber \\ & & 
 +  \sum_{l} \sqrt{\kappa_l} 
 %\frac{\partial O_l^*}{\partial c_j^*} 
 \frac{\partial O_{l}^*}{\partial{c^*_m}}    dZ_{l} 
 \hspace{5mm}
\end{equation}
where $d{Z} = (dX + i dY)/\sqrt{2}$ is a complex Wiener increment, in which $dX$ and $dY$ are independent Wiener increments, i.e., zero-mean normal deviates with variance $dt$.  
Using the jump operators  ${O}_{m}^{(1b)}$, ${O}_{mno}^{(3b)}$ defined above, and adding their noise contributions in quadrature, we find
\begin{eqnarray}
\label{eq:SDEbcOneAndThreeBody}
 d\bc  &=&  \left[ \frac{2 g}{i\hbar}  \sum_\alpha (\bc^\dagger f_\alpha \bc) f_\alpha \bc +\frac{q}{i\hbar} f_z^2 \bc 
+ A 
  \right] d t
 %  \nonumber \\ & & 
 + {\bf B}^{(\bc)} \cdot d{\bf Z}  \hspace{5mm}
\end{eqnarray}
where $d{\bf Z}$ is a vector of three complex Wiener increments as defined above and 
\begin{eqnarray}
\label{eq:ABOneAndThreeBody}
A &=& -  \frac{\rate{1}}{2} \bc - \frac{\Gamma_3}{2} |\bc|^{8/5} \bc \\
({B}^{(\bc)}_j)^2 &=&  \frac{\Gamma_{1}}{2}  +  \frac{5 \Gamma_3}{8} |\bc|^{-2/5} \left( |\bc|^{2} + \frac{23}{25} |c_j|^{2} \right).
\end{eqnarray}
%{B}^{(\bc)}_j &=& \sqrt{ \frac{k_{1} + \Gamma_3 |\bc|^2( |\bc|^2 + 2 |c_j|^2) }{2}}

We use fourth-order Runge-Kutta explicit integration \citeSI{WernerJCP1997} to evaluate the trajectories.  
Statistics, e.g., $\langle F_x \rangle$ or ${\rm var}(F_y)$,  are computed as the corresponding population statistic on the set of evolved values, e.g., ${\rm mean}\{ {\bf c}^*_i f_x {\bf c}_i\} $ or ${\rm var} \{ {\bf c}^*_i f_y {\bf c}_i \}$. Because the calculation is run in a frame rotating at the Larmor frequency, the observed results are scattered about the ideal value $F_y = 0$, and the atomic contribution to the angular mean squared error is simply $\langle \delta \theta^2 \rangle_F = \langle F_y^2 \rangle/\langle F_x \rangle^2$. [[mwm: the final $^2$ has been added 211202.]]

\section{Readout noise} \label{sec:Readout} \renewcommand{\Tevol}{T}
We experimentally prepare SBECs of \rb atoms in the $f=1, m = +1$ ground state under a bias field along direction $z$ and strength $B=\SI{29}{\micro T}$, which induces Larmor precession at angular frequency $\omega_L = 2 \pi \times \SI{200}{\kilo\hertz}$. A radio-frequency $\pi/2$ pulse is applied to tip the spins to the $xy$ plane. After a free evolution time $\Tevol$ we detect the spin precession by Faraday rotation, sending 60 pulses each of \SI{200}{ns} duration containing \SI{2e6}{photons} to observe rotation angles $\varphi_i$  at times $t_i$, $i=1,\ldots ,60$. Representative data are shown in \autoref{fig:SetupSequence} {\bf d} and are well described as a free-induction-decay signal. We parametrize the signal plus noise as
\begin{eqnarray}
\label{eq:dynamics}
\varphi_i &=& G_1 [\cos(\omega_L \tProbe_i){F}_y{(\Tevol)} +\sin(\omega_L \tProbe_i){F}_x{(\Tevol)}] e^{-\tProbe_i/\Trelax} 
\nonumber \\ &  &  
+ \varphi_i^{(\rm RO)}
\end{eqnarray}
where $G_1$ is the effective atom-light coupling in radians per spin, $\tProbe_i \equiv t_i - \Tevol$ is the time since the start of probing, 
%$\hat{z}$ is a unit vector along the probing direction, $R_x[\phi]$ indicates rotation about $x$ by angle $\phi$,  
${\bf F}(\Tevol)$ is the collective spin at the start of probing, 
 $1/\Trelax$ is the spin-relaxation rate due to probe scattering and $\varphi_i^{(\rm RO)}$ is the readout noise.  $G_1=\SI{2.5+-0.1e-7}{rad/atoms}$ is found by fully polarizing the atoms along $y$, such that $F_y = N$, and measuring $\varphi$ by Faraday rotation. $N$ is then measured by absorption imaging.  $\Trelax = \SI{29.7}{\micro \second}$, found by fitting {free-induction decay}s as in \autoref{fig:SetupSequence} {\bf d}.

To determine the atomic precession angle from a {free-induction decay} we define the angle estimator $\hat\theta_{e} \equiv \arctan[\hat{F}_{x}(T), \hat{F}_{y}(T)]$ in terms of the parameters $\hat{F}_{x}(T), \, \hat{F}_{y}(T)$ that make the best least-squares fit of \autoref{eq:dynamics} to a given {free-induction-decay} $\{\varphi_i\}$ with the previously determined $G_1$ and $\Trelax$. 
By propagation of errors, and due to the fit function's linear dependence on $F_x$ and $F_y$, the estimator's mean squared error is 
\begin{eqnarray}
%{\rm MSE}(\hat{\theta}) = \frac{{\bf r} \cdot \Gamma^{({\rm RO})} \cdot {\bf r}}{\langle {F}_z^2(\Tevol) +  {F}_y^2(\Tevol) \rangle} + \var (\theta)_{\rm PN},
%{\rm MSE}(\hat{\theta}) 
\langle  \delta \theta^2 \rangle_{\rm RO} &=& \frac{{\bf r^T} \cdot \Gamma^{({\rm RO})} \cdot {\bf r}}{N_0^2 \exp[-2\Tevol/T_2]} 
\label{eq:vartheta}
\end{eqnarray}
where ${\bf r} \equiv ( \cos{\theta}, -\sin{\theta})^T$ is a projector on the azimuthal direction, and $\Gamma^{({\rm RO})}_{ij}$ is the covariance matrix of the contribution made by $\varphi_i^{(\rm RO)}$ to the fit parameters. 

To evaluate \autoref{eq:vartheta}, we note that $\Gamma^{({\rm RO})}$ can be directly measured:  we collect 40 traces $\{\varphi_i\}$ at time $T$ with no atoms in the trap. We then fit \autoref{eq:dynamics} using the $G_1$ and $\Trelax$ obtained previously. The result is
{
\begin{eqnarray}
\Gamma^{({\rm RO})}&=&\left[  \left( \begin{array}{ccccc}
184&  -2 \\
-2 & 222
\end{array} \right)\pm \left( \begin{array}{ccccc}
38 & 30 \\
30 & 46
\end{array} \right) \right] \times 10^3. \hspace{9mm}
\end{eqnarray}
}
%\begin{eqnarray}
%\Gamma^{({\rm RO})}&=&\left[  \left( \begin{array}{ccccc}
%18.4&  -0.2 \\
%-0.2 & 22.2
%\end{array} \right)\pm \left( \begin{array}{ccccc}
%3.8 & 3.0 \\
%3.0 & 4.6
%\end{array} \right) \right] \times 10^4. \hspace{9mm}
%\end{eqnarray}
Combining the above, we find the readout noise reaches its minimum value of $\langle \delta {\theta} ^2\rangle_ {\rm RO}=\SI{1.08+-0.24e-4}{rad^2}$ when $\Tevol=\Ttwo/2$.

\section{SMA validity conditions}
\label{sec:SMAconditions}

Two criteria for the validity of the SMA are found in the literature for the scenario of interest, in which a $F=1$ condensate precesses about an orthogonal magnetic field. The first compares the ferromagnetic energy associated with a spatial overlap of the different $m_F$ states to the kinetic energy associated with a domain wall, to derive the condition $\rTF \ll  \xi_s \equiv 2 \pi \hbar/\sqrt{2 M |g_2| n}$, where $\xi_s$ is known as the spin-healing length \citeSI{HoPRL1998, Law1998SMA}. The second criterion derives from a consideration of dynamical stability \citeSI{makela2011}: In a plane wave scenario, spin-wave perturbations to an initially uniform spin precessing at $\omega_L=p/\hbar$ are non-increasing for wavelengths smaller than $\lambda_{\rm min} = 2 \pi \hbar/\sqrt{2 M ( |g_2| n-g_2 n+ q)}$.  A second condition for the SMA is then $\rTF \ll \lambda_{\rm min}$.  We note that for ferromagnetic interactions ($g_2<0$), but not for antiferromagnetic ones, this second condition is stricter than the first, because $\lambda_{\rm min} < \xi_s$.

\section{Duty cycle}
\label{sec:dutycycle}
While the main result of this work is a single-shot sensitivity, i.e. the noise level when measuring a field over a continuous interval $T$, it is also interesting to consider averaging multiple sequential sensor readings to obtain a time-averaged estimate for the field.  In this multi-shot scenario, the dead time between measurements must be accounted for in the energy resolution per bandwidth.  Including the $\SI{30}{\second}$ required to produce the next SBEC sample we  find a multi-shot sensitivity of $\SI{344+-39}{\femto\tesla\per\sqrt\hertz}$, and an energy resolution of $\langle \delta B^2 \rangle VT/(2\mu_0) = \SI{0.48+-0.11}{\hbar}$, which is also significantly below  $\hbar$ and well below any previously reported value. 

\section{Data availability}
{Data and data analysis codes are available for download at \cite{PalaciosZ2021art}.}

}

\acknowledgements
{We thank Luca Tagliacozzo for insightful feedback.  Work supported by 
H2020 Future and Emerging Technologies Quantum Technologies Flagship projects MACQSIMAL (Grant Agreement No. 820393) and  QRANGE (Grant Agreement No.  820405); 
H2020 Marie Sk{\l}odowska-Curie Actions project ITN ZULF-NMR  (Grant Agreement No. 766402); 
Spanish Ministry of Science  ``Severo Ochoa'' Center of Excellence CEX2019-000910-S, and project OCARINA (PGC2018-097056-B-I00 project funded by MCIN/ AEI /10.13039/501100011033/ FEDER ``A way to make Europe''); Generalitat de Catalunya through the CERCA program; 
Ag\`{e}ncia de Gesti\'{o} d'Ajuts Universitaris i de Recerca Grant No. 2017-SGR-1354;  Secretaria d'Universitats i Recerca del Departament d'Empresa i Coneixement de la Generalitat de Catalunya, co-funded by the European Union Regional Development Fund within the ERDF Operational Program of Catalunya (project QuantumCat, ref. 001-P-001644); Fundaci\'{o} Privada Cellex; Fundaci\'{o} Mir-Puig;
17FUN03 USOQS, which has received funding from the EMPIR programme co-financed by the Participating States and from the European Union's Horizon 2020 research and innovation programme. CONACYT 255573 (M\'exico) PAPIIT-IN105217 (UNAM)}

%H2020 Future and Emerging Technologies Quantum Technologies Flagship projects MACQSIMAL (Grant Agreement No. 820393) and  QRANGE (Grant Agreement No.  820405); 
%H2020 Marie Sk\l{}odowska-Curie Actions project ITN ZULF-NMR  (Grant Agreement No. 766402); 
%Spanish MINECO projects OCARINA (Grant No. PGC2018-097056-B-I00), and the Severo Ochoa program (Grant No. SEV-2015-0522); 
%Generalitat de Catalunya through the CERCA program and RIS3CAT project QuantumCAT; 
%Ag\`{e}ncia de Gesti\'{o} d'Ajuts Universitaris i de Recerca Grant No. 2017-SGR-1354; 
%Fundaci\'{o} Privada Cellex; Fundaci\'{o} Mir-Puig; EMPIR project 
%17FUN03-USOQS; CONACYT 255573 (M\'exico) PAPIIT-IN105217 (UNAM). }

\showmethods{} % Display the Materials and Methods section

%\renewcommand{\thesection}{}
%\section*{References}
\bibliographystyle{./apsrev4-1no-url}
\bibliography{./SBECMAG.bib}%,../../CommonLatexFiles/MegaBib}

\end{document}